\documentclass{aa} 
\usepackage{graphics}

\def\mincir{\raise -2.truept\hbox{\rlap{\hbox{$\sim$}}\raise5.truept 
\hbox{$<$}\ }} 
\def\mincireq 
{\hbox{\raise0.5ex\hbox{$<\lower1.06ex\hbox{$\kern-1.07em{\sim}$}$}}} 
\def  \magcir{\raise -2.truept\hbox{\rlap{\hbox{$\sim$}}\raise5.truept 
\hbox{$>$}\ }}

\begin{document} 
 
\thesaurus{ 03(11.11.1;       
	    11.19.2;          
	    11.19.6)}         
 
\title{
Maximal halos in high-luminosity spiral galaxies 
}

\author{ 
	P.~Salucci\inst{1} and 
	M.~Persic\inst{2}} 
 
\offprints{P.~Salucci; e-mail:{\tt salucci@sissa.it}} 
 
\institute{ 
SISSA, via Beirut 2-4, 34013 Trieste, Italy 
	\and  
Trieste Astronomical Observatory, via G.B.Tiepolo 11,  
	34131 Trieste, Italy} 
\date{Received..................; accepted...................} 
 
\maketitle 
\markboth{Salucci \& Persic: Maximal halos in galaxies}{} 
 
\begin{abstract} 
We test the  halo dominated mass model, recently proposed for high-luminosity  
galaxies by Courteau and Rix (1998), against rotation velocity data for a large 
sample of $L \sim L_*$ spirals. We find that this model does not reproduce the 
general kinematic properties required. The alternative disc dominated model, on 
the contrary, easily fits the data. 
\end{abstract} 
 

\section{Introduction} 
 
In a recent paper, Courteau \& Rix (1998; hereafter CR98) have used the residuals 
of the well known relationships of spiral galaxies, luminosity {\it vs.} velocity 
and luminosity {\it vs.} radius, to estimate the disc-to-halo mass ratio $\beta$ 
at $R =2.2 \,R_D$\footnote {The radius where the rotation curve of a thin 
exponential disc of lenghtscale $R_D$ peaks (Freeman 1970).}. Hereafter, 
$V_{2.2} \equiv V(2.2\, R_D)$ and $R_{opt} \equiv 3.2 R_D$. The argument of CR98 
relies on the idea that, for a given mass, more compact self-gravitating stellar 
discs have higher rotation speeds than less compact ones (according to a $R_D^{-1/2}$ 
scaling). This effect should be observed in the rotation curves, if the disc component 
is dynamically important, and should disappear if the halo is dominant.  
More specifically, the effect can be measured by correlating the deviation 
$\partial {\rm log}V_{2.2}$ that an  object has with respect to the mean  
velocity--luminosity relation, with the deviation $\partial {\rm log}R_D$ 
that the object has with respect to the mean radius--luminosity relation. 
For disc dominated systems, one should expect $\partial {\rm log}V_{2.2} / 
\partial {\rm log}R_D =-0.5$, while for halo dominated ones it is likely that 
$\partial {\rm log}V_{2.2} / \partial {\rm log}R_D$ $\mincir 0$. In large samples 
of $\sim L_*$ spirals, CR98 find an ensemble value of $\partial {\rm log}V_{2.2} / 
\partial {\rm log}R_D \mincir 0$, which is interpreted as evidence that, on 
average, the disc contribution to  $V_{2.2}$ is far from maximal (see below). 
The actual value of $\beta$ is estimated by computing the final halo mass 
distribution by adiabatically contracting (Blumenthal et al. 1986) a proto-disc 
inside an initial standard CDM halo (e.g., Navarro et al. 1996) into a final 
configuration with $R_D \simeq 3$ kpc. So, CR98 find 
$\beta=0.36^{+0.13}_{-0.11}$ for galaxies with $0.3 \mincir L_B/L_* \mincir 3$ 
\footnote{log$\,(L_*/L_\odot)=10.2$; see Rhee (1996) for the $B$-magnitudes of 
the CR98 sample. We adopt H$_0=75$ km s$^{-1}$ Mpc$^{-1}$. No result of this paper 
depends on this value.} (virtually independent of the initial DM distribution or the 
exact disc scale length or the presence of a [realistic] stellar bulge).

Before discussing this result, let us comment on the procedure adopted by CR98. Three 
crucial assumptions have been made: {\it 1)} the residuals of both relationships 
are essentially due to differences in the mass structure between each individual galaxy 
and the "average" galaxy; {\it 2)} the residuals are not affected by discrepancies 
between redshift distancies and true distances, observational errors and selection 
bias; {\it 3)} the two relationships are intrinsically log-linear and/or do not have 
hidden parameters. The validity of these assumptions is far from certain. Moreover, 
the method, devised to determine the average value of $\beta$ in a sample of galaxies, 
is poorly suited to detect systematic and/or random variations of $\beta$ within the 
sample itself. While these issues do not affect the main thrust of the CR98 idea, it is 
likely that they strongly modify the interpretation.   

Independently of the above issues (to be considered in a separate paper), the CR98 
claim of $\beta \simeq 0.36$ in {\it high-$L$} spirals is quite surprising. 
(Much less so if referred to {\it low-$L$} spirals, e.g. Persic \& Salucci 1990.)
Direct modelling of rotation curves (RCs) of high-$L$ galaxies, based on the maximum 
disc hypothesis (van Albada \& Sancisi 1986; Sancisi \& van Albada 1987), is able to 
successfully fit the RCs out to $\mincir 2\, R_D$ with  $\beta \sim 1$ (see: Kalnajs 
1983; van Albada et al. 1985; Kent 1986; Begeman 1987; Broeils 1992; Broeils \& Courteau 
1997). Best-fitting multicomponent mass modelling, applied to extended high resolution 
RCs, has also found disc-dominated solutions for high-$L$ galaxies (e.g.: Broeils 1992; 
see references in section 4.) The RC profile methods, that are powerful 
diagnostics in quantifying the disc-to-total mass discrepancy, consistently find that 
$\magcir L_*$ galaxies as disc dominated, i.e. $\beta \sim 1$ (Persic \& Salucci 1988, 
1990; Salucci \& Frenk 1989; Casertano \& van Gorkom 1991; Salucci \& Persic 1997). 
Finally, the swing amplifier constraints, coupled with stellar population arguments, 
seem to favour (near-)maximum discs (see Athanassoula et al. 1987 and Bosma 1998).
 
Despite this, CR98 are not alone in their claim of halo-dominated (HD) high-$L$ spirals. 
Mass models, built from the properties of the stellar velocity dispersion, find low values 
of $\beta$ (Bottema 1997); the swing amplifier constraints may also imply subdominant 
discs (Fuchs et al. 1998). Moreover, the local Galactic disc column density of $(50-70) 
M_\odot$ pc$^{-2}$, estimated by studying the {\it z}-motions  of local stars as a function 
of their height above the Galactic plane, is interpreted as evidence of a dominant halo 
(e.g., Kuijken \& Gilmore 1991 and Kuijken 1995; but see Sackett 1997 and Dehnen \& Binney 
1997). 
 
On the cosmological side, the universal CDM density profile (Navarro et al. 1996, 1997) 
features only a small central region where the halo rotation curve $V_h(R)$ increases 
with radius, while it has $V_h(R) \approx const$ for $R>R_D$ that, combined with 
$V(R) \simeq const$ observed in $\sim L_*$ galaxies, implies a dominant halo.
 
\begin{figure} 
\resizebox{\hsize}{!}{\includegraphics{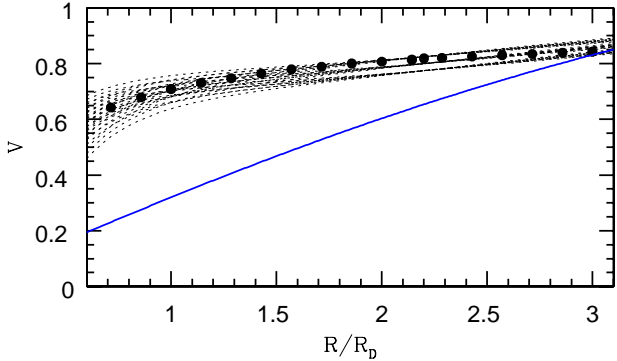}} 
\caption{ 
The halo velocity profiles of generic HD mass models with different values of
$\beta <0.5$ which reproduce high-$L$ RCs (dotted lines). Also shown is the
CR98 halo mass model (filled circles). The solid line is the PSS halo RC for 
high-$L$ galaxies, plotted to explicit the difference between the HD and DD 
halo RCs. Velocities are in units of $V_{2.2}$, radii are in units of $R_D$. 
} 
\label{fig:......} 
\end{figure} 
 
The case of HD mass models for high-$L$ spirals is in (indirect) disagreement 
with our previous work (Persic \& Salucci 1990; Persic, Salucci \& Stel, 1996 [PSS]). 
However, in view of the strong CR98 claim, it is worthwhile to investigate the issue 
further by means of specific tests.
 
Notice that, for high-$L$ galaxies, whose RCs are well described 
by $V(x) \propto x^c - A\, x^{-\alpha}$, with $x \equiv R/2.2 \, R_D$, $c \simeq 0$, 
$A \simeq 0.1$, $\alpha \simeq 1$, the CR98 mass model represents the whole class of 
HD models. In fact, a property of the above rotation curve is that  any HD model (i.e., 
with $\beta < 0.5$) will have a halo velocity profile, $V_h(R)=\sqrt{(V^2-V^2_{disc})}$, 
very similar to the one adopted  by CR98, independent of $\beta$ and independent also 
of the precise values of $c,\ A, \ \alpha$ (see Fig. 1) in high-$L$ objects. Then, a 
test performed on the CR98 model applies also to any HD mass model of high-luminosity spirals. 
 
In detail, in this paper we will devise two tests that, in the spirit of Salucci \& Frenk 
(1989), single out two clear kinematical properties of RCs, and we will then check whether 
these are compatible with the CR98 model; failing that, we will iterate the tests on a disc 
dominated (DD) model. 
 
The plan is the following: in section 2 we determine the synthetic RC of high-$L$ spirals; 
in sections 3 and 4 we perform two tests concerning the inner and outer rotation curves; in 
section 5 we discuss the results obtained and their implications.

\section{Rotation Curves of High-L Spirals} 
 
In this section we coadd individual spiral RCs into a synthetic curve. This 
procedure, pioneered by Rubin et al. (1985) separately for Sa, Sb and Sc galaxies, was 
extensively used by PSS. In particular, for a sample of $\sim$1000 optical RCs of mostly 
late spirals, PSS have shown that: (1) within each given luminosity interval, the 
scatter of RCs around their mean is comparable with the scatter of data points around a 
given individual curve (i.e., the cosmic scatter is comparable with the error); and (2) 
the sequence of galaxy luminosities determines a sequence of RC shapes. [A Principal 
Component analysis of the same sample (Rhee 1996; see also Roscoe 1998) confirms the 
above results.] Moreover, at a given luminosity, there is practically no difference 
between the {\it individual} profiles of (high quality) RCs and the corresponding 
{\it coadded} profile (PSS). This evidence indicates that, as far 
as {\it the gross features of their mass structure} are concerned, spirals within a 
suitably small interval of luminosity can be considered as essentially alike: 
the coaddition procedure, therefore, eliminates most of the observational errors, 
extracting the relevant information stored in individual curves. 
 
CR98 use two samples of galaxies. One comprises over a hundred $\sim L_*$ Sb/Sc galaxies 
from Courteau (1992); the other comprises several hundred late spirals from the Mathewson 
et al. (1992; hereafter MFB) sample with ${\rm M}_I <-20.25$. We then select from Persic 
\& Salucci (1995; hereafter PS95) the RCs of MFB objects in the same luminosity range as 
CR98. From this subsample we will generate a synthetic curve by coadding the individual 
curves. The procedure used is the following: we select galaxies with  ${\rm M}_I <-20.25$ 
and $i < 85^{\circ}$ (the RC of objects of higher inclination are excluded because, due to 
obscuration effects, the H$\alpha$ position--velocity curves may not correspond 
to true rotation velocities: see Bosma et al. 1992 and Bosma 1995). We then find 
554 objects, with: $<{\rm M}_I>= -21.71$, $<R_D>=3.6$ kpc. \footnote{ The average surface 
luminosity profile of the objects in the sample is well represented by a thin exponential 
disc (PSS).}  
 
Following PSS we coadd the (normalized) velocity data and obtain a raw synthetic RC 
with about 12000 measurements, $V({R \over R_{opt}}; ~ 0.3 \mincir {L \over L_*}  
\mincir 3$).\footnote{ In the following: $r \equiv R/R_{opt}$.} We then form the smoothed 
synthetic RC of $\sim L_*$ galaxies, $V(r)$, by averaging the velocity data over 24 radial 
bins of size $0.05\, R_{opt}$ out to $1.2\, R_{opt}$ (see Fig.2). {\it $V(r)$ is the average RC 
of} (essentially) {\it the same sample whose residuals have been studied by CR98.}

\section{Test I. Mass Modelling of Inner Rotation Curves} 
 
\begin{figure} 
\resizebox{\hsize}{!}{\includegraphics{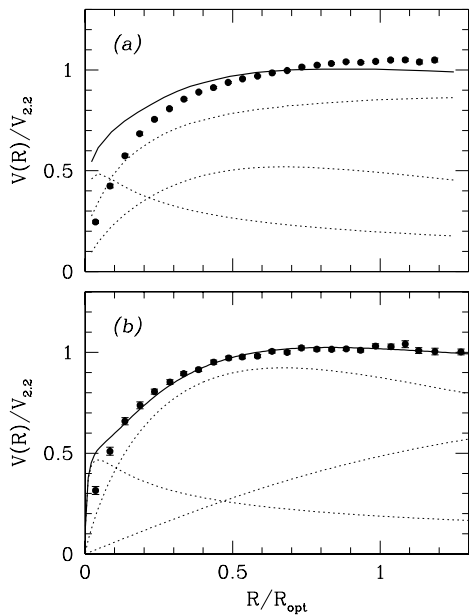}} 
\caption{ 
{\it (a)} Synthetic RC of $\sim L_*$ galaxies (filled circles), with superposed the halo
dominated CR98 model (solid line). 
{\it (b)} The disc dominated mass model (solid line) superposed on the synthetic RC of 
high-$L$ galaxies (filled circles). 
In both panels, velocities are in units of $V_{2.2}$ and radii are in units of $R_{opt}$; 
the bulge, disc and halo velocity components are denoted by dotted lines). 
} 
\end{figure} 
 
In this section we check whether the innermost (i.e., for $R  \mincir 2\, R_D$) RCs of 
$\sim L_*$ spiral galaxies are halo dominated. The test is straigthforward because for 
$R \sim R_D$ the velocity profile of a Freeman (1970) disc is completely different from 
that of a post-infall CDM halo.
 
We first fit $V(r)$ with the HD mass model which has allowed CR98 to model the residuals 
$\partial {\rm log}V_{2.2}$ and $\partial {\rm log}R_D$ (see Fig.7 of CR98) for the same 
sample that we have used to construct $V(r)$. In detail, the model includes: 
{\it (i)} an exponential disc with $R_D= 3.5$ kpc;  
{\it (ii)} a stellar bulge with velocity profile peaking at $R_m/R_{opt}=0.045$; 
{\it (iii)} a post-infall CDM dark halo (see Navarro 1998). For this mass distribution 
$V_h(r)$ has an initial $r^{1/2}$ profile and then a rapid convergence to a very shallow 
slope: setting $V_h(r) \propto r^{k(r)}$, then: $k(0.15) \simeq 0.25$ and $k \simeq 0.05$ 
for $r > 0.7$.  

The CR98 model is plotted in Fig.2{\it a} alongside its individual components. It has one 
distinctive 
feature: about $70\%$ of the maximum velocity is reached at $r = 0.10$. By comparison, the 
exponential thin disc reaches the same fraction of the maximum velocity at a much larger 
radius, $r=0.21$. Notice that the step-like shape of the halo dominated RC is not due the 
bulge, which reaches its maximum velocity at the much smaller radius $r=0.05$, but just 
reflects the shape of $V_h(r)$ that, for $r < 0.4$, is steeper than $V_{disc}(r)$. As a 
result, we realize that the model is inconsistent with the data, by predicting, for $0<r 
\mincir 0.4$, a curvature in $V(r)$ very different from the observed one. 

This failure does not occur only when testing the coadded RC, but also when we check 
individual RCs of high luminosity spirals that result severely inconsistent with the curve 
generated by the gravitational field of a dominant $r^{-2}$-halo [see the cases of NGC 2179 
and NGC 2775 (Corsini et al. 1998), of NGC 5055 (Thornley \& Mundy 1997) and of NGC 5033 
(Thean et al. 1997), and a few more cases in Begeman (1987) and Rhee (1996)]. 
 
Next, we fit $V(r)$ with a DD model. This should not be considered as a further test, but 
rather as the necessary check that the coadded RC can be reproduced by a 'natural' mass 
distribution, i.e. the one which follows the distribution of light. Applying DD models to 
samples with large luminosity ranges (as in the case of CR98) requires, however, the 
luminosity bins to be smaller. Then, in this case we analyse the RCs of MFB objects (see 
PS95) having ${\rm M}_I <-21.80$. (The reader is referred to PSS for the analogous results 
concerning lower-$L$ galaxies.) The present subsample comprises 267 objects with global 
properties: $<{\rm M}_I>= -22.43$, $<{\rm log}\,L_B> = 10.52$, $<V_{2.2}> = 210$ km s$^{-1}$ 
and $<R_{D}>=4.3$ kpc.
 
The DD mass model we use includes:  
{\it (i)} a thin exponential disc with $R_D=4.3$ kpc;
{\it (ii)} a bulge, with a Hernquist velocity profile (practically 
indistinguishable from that used by CR98): 
\begin{equation} 
V_b(r) ~ =~ 106 ~ \sqrt{R \over R_m} ~
	{2 \over 1+R/ R_m}\,, 
\end{equation} 
where $R_m = 0.63$ kpc is the radius where the bulge velocity peaks;  
{\it (iii)} a dark halo with velocity profile: 
\begin{equation} 
     V_h^2(r) ~=~ V^2(1) ~ (1-\beta) ~ (1+a^2) ~ {r^2 \over r^2+a^2}\,, 
\end{equation} 
where $V(1) =215$ km s$^{-1}$, $a=5$ is the "velocity core radius" (in units of $2.2 
R_{D}$), and $\beta =0.85$ is essentially the stellar mass fraction at $2.2\,R_D$ 
(the bulge contributes only 6\% of the mass at this radius). Our adopted DD model, 
therefore, {\it is not a best-fit model}, but is chosen such as to only differ from 
the previously used HD model by (mainly) the disc dominance: the bulge is kept 
virtually frozen (in normalized units), and the halo velocity profile is a simple fit 
to the velocity residuals. The DD model easily reproduces the data for $r \magcir 0.1$ 
within their rms errors (see Fig.2{\it b}): at very small radii the detailed modelling 
of the bulge is crucial but is beyond our scope here.

Then we can conclude that, inside $\sim 2R_D$, the synthetic RC of high-$L$ objects 
does not trace the dark halo mass distribution: in fact, it does precisely trace the 
mass distribution of the stellar disc.
 

\section{Test II. The Inner vs. Outer RC Slope} 
 
A second test to check the ability of mass models to fit the RCs relies on the 
analysis of the outer RC profiles. The shapes of RCs can be essentially described 
by two quantities: the inner slope $\nabla \equiv d{\rm log}V/d{\rm log}R|_{R_{opt}}$ 
and the outer slope $\delta \equiv [V(2\,R_{opt}) - V(R_{opt})]/V(R_{opt})$ (see 
PSS). Used in combination, $\delta$ and $\nabla$ can help discriminating among 
apparently similar mass models. 

Let us consider, for example, two radically different models that lead to curves 
approximately flat at $R \simeq R_{opt}$: the "light disc + $r^{-2}$ halo" model, 
and the "dominant disc + constant-density halo" model. \footnote{The "light disc + 
constant-density halo" and "dominant disc + $r^{-2}$ halo" models lead to curves 
that are respectively {\it increasing} and {\it decreasing} at $R_{opt}$.} 
In the former case, $({V_{disc} \over  V})^2 \, {R \over V_{disc}}
{dV_{disc}\over dR}$ and $({V_h \over V})^2 \, {R \over V_h} {dV_h \over dR}$ 
are both small in absolute value ($< 0.1$) but are opposite in sign, thus 
$$
\nabla\,\,  \equiv \, \, {V^2_{disc}\over   V^2} \, {R \over V_{disc}}
{dV_{disc}\over dR} \, + \, {V^2_h \over V^2} \, {R \over V_h}
{dV_h \over dR} \,\, \simeq \,\, 0
$$ 
implies also $\delta \simeq 0$; moreover, since the disc contribution for $R > 
R_{opt}$ is negligible, then $\nabla \simeq \delta$ (specifically, the CR98 model 
has: $\nabla = -0.04$, $\delta = -0.04$). 
In the second  case, $({V_{disc} \over V})^2 \, {R \over V_{disc}} {dV_{disc} 
\over dR}$ and $({V_h \over V})^2 \, {R \over V_h} {dV_h \over dR}$ are both a 
factor 3-4 larger than in the previous case: they approximately compensate each 
other in the range ${1 \over 2} R_{opt} \sim R_{opt}$, but further out 
the disc term takes over because ${R \over V_{disc}} {dV_{disc} \over dR}$ 
increases with radius while ${R \over V_h} {dV_h \over dR}$ decreases: then  
the dominant disc model has $-0.1\mincir \nabla \mincir 0.1$ and $\delta 
< \nabla$, with $-0.2 \mincir \delta \mincir 0$.  
 
We now check these predictions versus the observations. From the literature we select, 
following PSS, 19 individual RCs of high-$L$ galaxies with a reliable profile out to 
$2\,R_{opt}$. These 19 curves are representative of the sample of 554 used for test I.
(See Figs.2 and 3 of PSS: the shape parameters $\nabla, \delta$ for the sample of 
individual RCs out of which these 19 are extracted, are consistent with the shape  
parameters of the coadded RCs built from essentially the same sample used in test I.)
Using these curves, in Fig.3 we see that very few objects have a pair 
$\delta, \nabla$ consistent with HD models (shaded area) and none lies in the region 
predicted by the CR98 mass  model; moreover, $\magcir 90\%$ of objects lay well inside 
the disc dominated region of the $\delta$-$\nabla$ plane. 

Finally, the constraint $\delta< \nabla < 0$, predicted by  the disc dominated model, 
is fully satisfied: the observed values are $<\delta>= -0.08\pm 0.007$ and $<\nabla>= 
-0.025 \pm 0.008$. Again, these values are inconsistent with the prediction of HD models:
$-0.04 \leq \nabla< \delta+\epsilon$, with $\epsilon =O(10^{-3})$. 
 
\begin{figure} 
\resizebox{\hsize}{!}{\includegraphics{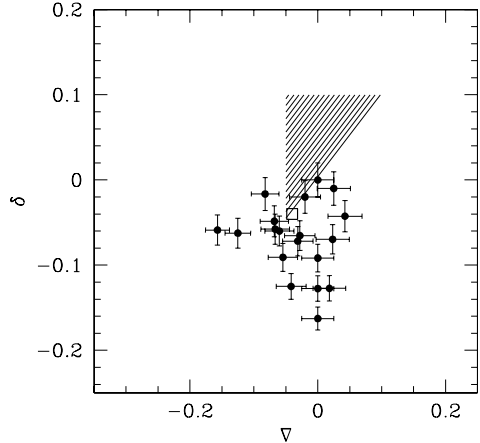}} 
\caption{
The distribution of high-$L$ galaxies in the inner/outer-slope ($\nabla, \delta$) plane. 
Data (filled circles) are compared with the CR98 model (the empty square). The shaded area 
represents the HD region.
} 
\end{figure}

\section{Discussion and conclusion} 
 
Could high-$L$ spirals be completely halo dominated? This can be effectively  
checked out by looking at their RCs: in fact, if the disc contribution to the 
RC is very small at all radii, the RC profile will essentially reflect that of 
the dark halo; otherwise, the stellar disc component will leave a recognizable 
imprint on the shape of the RC.  
 
By analysing the data for 554 inner RCs and 19 outer RCs of $\sim L_*$ spirals, we 
have uncovered two clear dynamical imprints of a dominant disc. The first consists 
in a shallow extended rise in the RC between $0.3 \,R_D$ and $1.3 \, R_D$, that 
results coincident with that predicted by the Freeman (1970) disc; the second consists 
in a net decrease, $(10 - 20)$ km s$^{-1}$ (i.e. $\sim 10\%$), of the circular 
velocity between one and two optical radii, again as predicted by a dominant disc.  
These features, predicted well before statistically relevant data were available 
(Salucci \& Frenk 1989), are inconsistent with the predictions of HD models (including 
the CR98 one). It is then obvious that mass models, featuring light discs and essentially 
constant velocity halos, give a poor fit to the data, while models featuring massive 
discs and DM halos with solid-body-like profiles out to $R\simeq R_{opt}$ match the data.  
    
On the basis of the results of two tests, one performed on the {\it synthetic} RC 
of high-$L$ galaxies (seconded by literature results on several individual RCs) and the 
other performed on about twenty {\it individual} RCs, we then reject the CR98 claim for 
halo dominated high-$L$ galaxies. \footnote{One further, non-dynamical argument. The 
CR98 model has $M_\star /L_B \simeq 1.0\,{ H_0\over 75\, {\rm km \,\, s}^{-1} {\rm 
Mpc}^{-1} }$ [with $M_\star$ the mass in stars (disc plus bulge)]: this, however, seems 
rather low for objects having $B-V \sim 0.7$ (e.g.: Tinsley 1981; Jablonka \& Arimoto 
1992) like high-$L$ spirals.}

We conclude by emphasizing the existence of a characteristic length-scale $R_L(M_b)$ 
such that, for radii $R < R_L$ the baryonic matter (with total mass $M_b$) accounts 
for most (e.g., $>80\%$) of the gravitating mass inside $R$, and for $R > R_L$ the dark 
component (whose contribution is minor at inner radii) rapidly becomes a major mass 
component. From PSS we can estimate: $R_L \sim 2\,R_D ({M_b\over 10^{11}M_\odot})^k$, 
with $k \sim 1/4$. For high-$L$ spirals, such "inner baryon dominance" extends to most 
of the disc region. 

\acknowledgements{We thank St\'ephane Courteau for useful exchanges. We also thank 
the referee, Albert Bosma, for several stimulating comments that have significantly 
improved the paper.} 
\vglue 0.5truecm

\def\ref{\par\noindent\hangindent 20pt} 
 
\noindent 
{\bf References} 
\vglue 0.2truecm 
 
\ref{Athanassoula, E., Bosma, A., \& Papaioannou, S. 1987, A\&A, 179, 23} 
\ref{Begeman, K. 1987, PhD thesis, University of Groningen} 
\ref{Blumenthal, G.R., Faber, S.M., Flores, R., \& Primack, J.P. 1986,  
	ApJ, 301, 27} 
\ref{Bosma, A., 1995, in NATO Advanced Reasearch Workshop ``The 
     Opacity of Spiral Disks'', eds. J.I.Davies \& D.Burstein 
     (Cambridge University Press), 317} 
\ref{Bosma, A., 1998, astro-ph/9812013} 
\ref{Bosma, A., Byun, Y.I., Freeman, K.C., \& Athanassoula, E. 1992,  
	ApJ, 400, L23} 
\ref{Bottema, R. 1997, A\&A, 328, 517} 
\ref{Broeils, A.H. 1992, PhD thesis, Groningen U.} 
\ref{Broeils, A.H., \& Courteau, S. 1997, in ``Dark and Visible Matter in 
	Galaxies'', ed. M.Persic \& P.Salucci, ASP Conference 117 (S.Francisco: 
	ASP), 74}   
\ref{Casertano, S., \& van Gorkom, J.H. 1991, AJ, 101, 1231} 
\ref{Corsini, E.M., et al. 1998, A\&A in press, astro-ph/9809366} 
\ref{Courteau, S. 1992, PhD thesis, Univ. of California, Santa Cruz} 
\ref{Courteau, S., \& Rix, H.-W. 1998, astro-ph/9707290 v2 (CR98)} 
\ref{Dehnen, W., \& Binney, J. 1998, MNRAS, 294, 429} 
\ref{Freeman, K.C. 1970, ApJ, 160, 811} 
\ref{Fuchs, B., Moellenhoff, C., \& Heidt, J. 1998, astro-ph/9806117}
\ref{Jablonka, P., \& Arimoto, N. 1992, A\&A 255, 63} 
\ref{Kalnajs, A.J. 1983, in IAU Symp. 100, ed. E. Athanassoula (Dordrecht: 
	Reidel), 87} 
\ref{Kent, S.M. 1986, AJ, 91, 1301} 
\ref{Kuijken, K., 1995, in IAU Symp. 164, "Stellar Populations"  
	(Dordrecht: Kluwer), 195} 
\ref{Kuijken, K., \& Gilmore, G. 1991, ApJ, 367, L9} 
\ref{Mathewson, D.S., Ford, V.L., \& Buchhorn, M. 1992, ApJS, 81, 413} 
\ref{Navarro, J.F. 1998, astro-ph/9807084} 
\ref{Navarro, J.C., Frenk, C.S., \& White, S.D.M. 1996, ApJ, 462, 563} 
\ref{Navarro, J.C., Frenk, C.S., \& White, S.D.M. 1997, ApJ, 490, 493} 
\ref{Persic, M., \& Salucci, P. 1988, MNRAS, 234, 131} 
\ref{Persic, M., \& Salucci, P. 1990, MNRAS, 245, 577} 
\ref{Persic, M., \& Salucci, P. 1995, ApJS, 99, 506 (PS95)} 
\ref{Persic, M., Salucci, P., \& Stel, F. 1996, MNRAS, 281, 27 (PSS)} 
\ref{Rhee, M.-H. 1996, PhD thesis, Groningen U.}  
\ref{Roscoe, D.F. 1998, A\&A, in press} 
\ref{Rubin, V.C., Burstein, D., Ford, W.K., Jr., \& Thonnard, N. 1985,  
	ApJ, 289, 81} 
\ref{Sackett, P.D. 1997, ApJ, 483, 103} 
\ref{Salucci, P., \& Frenk, C.S. 1989, MNRAS, 237, 247}  
\ref{Salucci, P., \& Persic, M. 1997, in ``Dark and Visible Matter in 
	Galaxies'', ed. M.Persic \& P.Salucci, ASP Conference 117 (S.Francisco: 
	ASP), 1} 
\ref{Sancisi, R., \& van Albada, T.S. 1987, in IAU Symp. 117 ``Dark Matter in 
	the Universe'', ed. J.Kormendy \& G.R.Knapp (Dordrecht: Reidel), 67} 
\ref{Thean, A.H.C., Mundell, C.G., Pedlar, A., \& Nicholson, R.A. 1997,  
	MNRAS, 290, 15} 
\ref{Thornley, M.D., \& Mundy, L.G. 1997, ApJ, 484, 202} 
\ref{Tinsley, B.M. 1981, MNRAS, 194, 63}
\ref{van Albada, T.S., Bahcall, J.N., Begeman, K., \& Sancisi, R. 1985,  
	ApJ, 295, 305} 
\ref{van Albada, \& T.S., Sancisi, R. 1986, Phil. Trans. Roy. Soc. A, 320, 447} 
\vfill\eject		
\end{document}